\documentclass[finalprint, 5p]{elsarticle}
\usepackage{graphicx}
\usepackage{amsmath}
\usepackage{makeidx}
\usepackage{amssymb}
\usepackage{float}

\begin{document}

\begin{frontmatter}

\title{Gross Pitaevskii Equation with a Morse potential: bound states and evolution of wave packets}
\author[label1]{Sukla Pal\corref{cor1}}
\ead{sukla@bose.res.in}
\author[label2]{Jayanta K. Bhattacharjee}
\address[label1]{Department of Theoretical Physics, S.N.Bose National Centre For Basic Sciences, JD-Block, Sector-III, Salt Lake City, Kolkata-700098, India}
\address[label2]{Harish-Chandra Research Institute, Chhatnag road, Jhunsi, Allahabad-211019, India}
\cortext[cor1]{Corresponding author}

\begin{abstract}
We consider systems governed by the Gross Pitaevskii equation (GPE) with the Morse potential $V(x)=D(e^{-2ax}-2e^{-ax})$ as the trapping potential. For positive values of the coupling constant $g$ of the cubic term in GPE, we find that the critical value $g_c$ beyond which there are no bound states scales as $D^{3/4}$ (for large $D$). Studying the quantum evolution of wave packets, we observe that for $g<g_c$,  the initial wave packet needs a critical momentum for the packet to escape from the potential. For $g>g_c$, on the otherhand, all initial wave packets escape from the potential and the dynamics is like that of a quantum free particle. For $g<0$, we find that there can be initial conditions for which the escaping wave packet can propagate with very little change in width i,e., it remains almost shape invariant.
\end{abstract}
\end{frontmatter}
\section{Introduction}
The wave packet dynamics of the usual Schrodinger equation has been extensively studied for the free particle as well as for different kinds of external potential. For the free particle, if a nonlinear term in the wave function is added to the Schr$\ddot{o}$dinger equation, then one has the non-linear Schr$\ddot{o}$dinger equation (NLSE)\cite{fg,n}. This too has been a subject of intense study and one has found various kinds of exact solutions like solitons, pulses, fronts etc.\cite{pp,1,2}. If one adds a potential, then one has the Gross Pitaevskii equation (GPE)\cite{a,b} which has been very useful for describing the process of Bose Einstein condensate (BEC)\cite{8,10} for a trapped gas (it should be noted that experimentally BEC has only been observed in a trapped gas). The trapping potential makes the problem far more difficult and the only case which has been reasonably well studied is the simple harmonic potential. In an experiment, it is actually far more convenient to tune the trapping potential and observe different behaviors of the condensate. With this in mind, we wanted to study the GPE with a Morse potential so that one can achieve a much greater flexibility in adjusting the potential. In fact, the potential can have bound states or no bound state at all, depending upon certain parameters of the potential. The corresponding dynamics is treated both for $g>0$ and $g<0$, where $g$ is the coupling constant of the non linear term.

The primary importance of GPE lies in the fact that it describes the dynamics of the condensate in the process of Bose Einstein Condensation (BEC)\cite{7,11}. The emergence of different kinds of soliton (dark, bright, grey \cite{c}-\cite{e} is well known in case of GPE. Also the controllable soliton emission \cite{f} has been investigated for GPE with a shallow trap and with negative interspecies interaction. In this article, we have dealt with the attractive and repulsive interatomic interaction separately for the Morse potential and the dynamics in both of these two cases have been explored both numerically and analytically.

Time dependent Gross Pitaevskii equation (GPE) with one spatial dimension has the following form:
\begin{eqnarray}
\label{eq1}
i\hbar\partial_t\psi =-\frac{\hbar^2}{2m}\nabla^2\psi+g|\psi|^2\psi+V_{ext}(x)\psi
\end{eqnarray} 
\section{Bound state}
\begin{figure}[H]
\includegraphics[angle=0,scale=0.6]{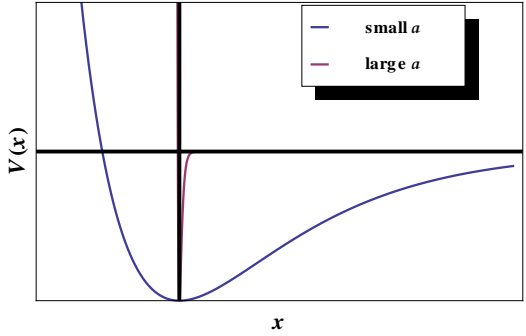}
\caption{Schematic diagram of Morse potential depicting the two extreme conditions. Blue curve shows approximately the oscillator limit ($a\ll 1$) and the violet one (for $a\rightarrow\infty$) indicates the free behavior for region of $x>0$.}
\label{fig1}
\end{figure}

The stationary states of Gross Pitaevskii equation (GPE) have the form $\psi(x,t)=e^{i\mu t/\hbar}u(x)$, so that $\mu u(x)=-\frac{\hbar^2}{2m}\nabla^2u+gu|u|^2+V_{ext}u$ and lowest value of $\mu$ is the ground state energy $E_0$ which is to be obtained by minimizing $E$ of Eq. (\ref{eq2}).
\begin{eqnarray}\label{eq2}
E[\psi(x)]=\int^{\infty}_{-\infty}[\frac{\hbar^2}{2m}|\nabla\psi|^2+V_{ext}|\psi|^2+\frac{g}{2}|\psi|^4]dx
\end{eqnarray}
 where, $V_{ext}(x)=D(e^{-2ax}-2e^{-ax})$. For $a<<1$, $V_ext(x)\sim x^2$ indicating the oscillator limit of the Morse. On the otherhand, for $a\rightarrow\infty$, $V_{ext}(0)=-D$ and $0$ else where for $x>0$ as is clearly depicted in Fig (\ref{fig1}).

 With the differential equation for $u$ given as
\begin{eqnarray}
\label{eq4}
-\frac{\hbar^2}{2m}\frac{d^2u}{dx^2}+D(e^{-2ax}-2e^{-ax})u+g|u|^u=\mu u
\end{eqnarray}
we can explore the asymptotic solution as $x\rightarrow-\infty$. The function $u$ has to vanish as $x\rightarrow\pm\infty$ being a bound state function and hence the dominant part of Eq. (\ref{eq4}) for $x\rightarrow-\infty$ will be $e^{-2ax}$ and we have in that range 
\begin{eqnarray}
\label{eq5}
-\frac{\hbar^2}{2m}\frac{d^2u}{dx^2}+De^{-2ax}u\sim 0
\end{eqnarray}
Making the substitution $e^{-ax}=y$, we find the exact asymptotic form of $u$ for $x\rightarrow-\infty$ ($y\rightarrow\infty$) to be of the following form
\begin{eqnarray}
\label{eq6}
u_{asy}=Ae^{-Ky}
\end{eqnarray}
where, $K^2=\frac{2mD}{\hbar^2a^2}$ and A is a constant to be determined from normalization in y-space. Since the ground state wave function will have no zeroes at any finite nonzero value of $y$ (note that $-\infty\le x\le\infty$ corresponds to $0\le y\le\infty$), we take $u(y)=\sqrt{N}y^{\alpha}e^{-Ky}$, to be the trial wave function for the minimization of $E$ given in Eq. (\ref{eq2}) and we obtain the following expression for the ground state energy,
\begin{eqnarray}
\label{eq7}
E(\alpha)=\frac{\hbar^2a^2}{2m}[\alpha^2+\alpha(1-2K)+\frac{agm}{\hbar^2a^2}\frac{\Gamma(4\alpha)}{2^{4\alpha}[\Gamma(2\alpha)]^2}]
\end{eqnarray}
If $g=0$ (i.e., usual Schr$\ddot{o}$dinger equation with a Morse potential), we get the critical $\alpha$ for minimization of energy is $\alpha=K-\frac{1}{2}$ and the ground state energy turns out to be  
\begin{eqnarray}
\label{eq8}
E=-\frac{\hbar^2a^2}{2m}\frac{1}{4}(1-2K)^2
\end{eqnarray}
which is the exact answer.It should be noted that for $K=\frac{1}{2}$, the binding energy reaches zero. i.e., for $K<\frac{1}{2}$ (equivalently for $a>\sqrt{\frac{8mD}{\hbar^2}}$, there can be no bound state in this potential. We now analyze the effect of non linear term. Returning to Eq. (\ref{eq7}), we consider the term involving `$g$' and and writing $\lambda=\frac{agm}{\hbar^2a^2}$, we get as the minimization condition 
\begin{eqnarray}
\label{eq9}
2\alpha+(1-2K)+\lambda\frac{d}{d\alpha}\frac{\Gamma(4\alpha)}{2^{4\alpha}[\Gamma(2\alpha)]^2}=0
\end{eqnarray}
Since the tightly bound situation correspond to large value of $K$, we use the above equation in the limit when $\alpha$ is reasonably large and $\Gamma(x)$ can be replaced by the asymptotic form 
\begin{eqnarray*}
\Gamma(x)\sim e^{-x}x^xx^{-1/2}\sqrt{2\pi}[1+\frac{1}{12x}+\cdots]
\end{eqnarray*}
We get from Eq. (\ref{eq9}) to the leading order, $\alpha$ to be given by
\begin{equation*}
\alpha=K-\frac{1}{2}-\frac{\lambda}{2\sqrt{2\pi}}\frac{1}{(2K-1)^{1/2}}
\end{equation*}
This gives a ground state energy of 
\begin{equation}\label{eq10}
E=\frac{\hbar^2a^2}{2m}[-\frac{1}{4}(2K-1)^2+\frac{\lambda}{2\sqrt{\pi}}(2K-1)^{1/2}]
\end{equation}
an equation which is accurate only for $K\gg1$. In Table 1, we show the comparison of ground state energies obtained from the actual solution of the minimization condition of Eq. (\ref{eq9}) and the energy obtained from Eq. (\ref{eq10}) for fixed values of $\lambda$ of $\lambda=1.0<\lambda_c$ and different values of $K$. As expected the approximation of Eq. (\ref{eq10}) gets closer to the answer obtained from  Eq. (\ref{eq9}) as $K$ increases. The increase of $\lambda$ weakens the ground state and for large $K$, we have the result that the critical $\lambda$ for disappearance of a bound state scales as $K^{3/2}$ i.e., $D^{3/4}$. 
\begin{table}[H]
 \centering 
\begin{tabular}{c  c  c  } 
\hline\hline 
$K$ & $E_{0}(from Eq. (\ref{eq10}))$ & $E_{0}(from Eq. (\ref{eq9}))$ \\ [0.5ex] 
\hline 
2 & -1.762 & -2.243\\
3 & -5.619 & -6.25\\
4 & -11.504 & -12.25\\
5 & -19.404 & -20.25\\ 
6 & -29.315 & -30.25\\[1ex]
\hline 
\end{tabular}
\caption{Results compairing the ground state energies from variational calculation (third column) with the expression (Eq. (\ref{eq10}) obtained considering the asymptotic series for different values of $K$ keeping $\lambda$ fixed at 1.0}
\end{table}
The interesting thing would be to explore the dynamics of an initial Gaussian wave packet in the region $\lambda>\lambda_c$ and $\lambda<\lambda_c$ for $\lambda>0$. In the next section we describethe wave packet dynamics both for $\lambda>0$ and $\lambda<0$. 

\section{Wave packet dynamics}
\begin{figure}[H]
\includegraphics[angle=0,scale=0.75]{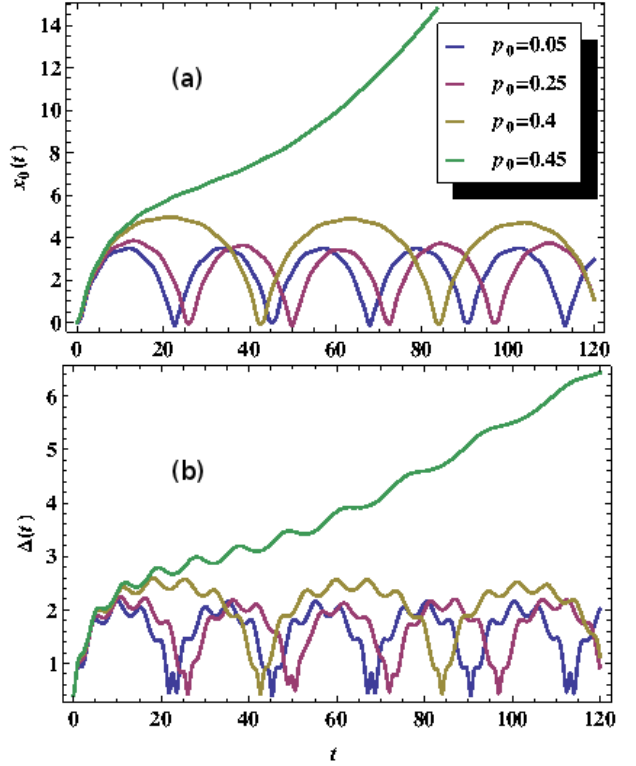}
\caption{Figure (a) shows the dynamics of the peak ($x_0$) of the wave packet for $\gamma(=0.5)<\gamma_c(=0.917)$ for four different values of $p_0$. The potential parameter $K=2$ and the initial $x_0=0$ while the width $\Delta_0=0.4$. For low values of $p_0$ the wave packet hits the potential barrier at right and reflects back and forth within the trap showing oscillatory behavior. With increase of $p_0$ the number of reflection reduces and finally at $p_0=p_{th}$ the wave packet comes out of the potential barrier which is represented by the linear increase of $x_0(t)$ at later time (green curve). The figure above shows that in this case $p_{th}=0.45$. Figure (b) shows the dynamics of the width ($\Delta$). All the parameters are same as are taken in figure (a). With lower values of $p_0$ when the wave packet hits the potential barrier and reflects back and forth within the trap, $\Delta(t)$ also shows nearly oscillatory behavior. With increase of $p_0$ the the wave packet finally escapes with increase of $\Delta(t)$ as $t$ increases (green curve).}
\label{fig2}
\end{figure}

The effective Hamiltonian of GPE with a Morse trap is given by
\begin{eqnarray*}
H=\frac{p^2}{2m}+D(e^{-2ax}-2e^{-ax})+g\int dx|\psi(x)|^4
\end{eqnarray*}
We make $H$ dimensionless by the following rescallings: $p=\sqrt{mD}\bar{p}$, $\bar{\psi}=\psi\sqrt{\frac{1}{a}}$, $x=\frac{\bar{x}}{a}$, $\Delta=\frac{\bar{\Delta}}{a}$, $t=\bar{t}\sqrt{\frac{a^2D}{m}}$, $\bar{V}=\frac{V}{D}$ and $\frac{g}{\sqrt{2\pi}}=\gamma\frac{D}{a}$. Hence the relation between $\lambda$ and $\gamma$ turns out to be $\lambda=\sqrt{2\pi}\gamma\frac{K^2}{2}$. Considering these transformations the dimensionless Hamiltonian turns out to be
\begin{eqnarray}\label{eq11}
\bar{H}=\frac{H}{D}=\frac{\bar{p}^2}{2}+(e^{-2\bar{x}}-2e^{-\bar{x}})+\sqrt{2\pi}\gamma\int d\bar{x}|\bar{\psi}(\bar{x})|^4
\end{eqnarray}
We consider Gaussian wave packets of the form given below
\begin{equation}\label{aa}
\psi(x,t)=\frac{1}{\pi^{1/4}\sqrt{\Delta(t)}}e^{-\frac{(x-x_0(t))^2}{2\Delta(t)^2}}e^{ip(t)x/\hbar}
\end{equation}
i.e., we assume an initially Gaussian form remains Gaussian with a changing centre and width. The initial shape corresponding to $\bar{x}_0=0$ and $\Delta=\Delta_0$ has the energy
\begin{eqnarray}\label{a}
\frac{\langle E_0\rangle}{D}=\frac{1}{2K^2\bar{\Delta}_0}+\frac{\bar{p_0}^2}{2}+[e^{\bar{\Delta}_0^2}-2e^{\frac{\bar{\Delta}_0^2}{4}}]+\frac{\gamma}{\bar{\Delta}_0}
\end{eqnarray}
The dynamics of the wave packet is governed by the following equations (we drop the bars with the understanding that all quantities are dimensionless).
\begin{eqnarray}\label{eq13}
\begin{aligned}
\frac{d\langle x\rangle}{dt} &=  \langle p\rangle\\
\frac{d\langle x^2\rangle}{dt} &= \langle xp+px\rangle\\
\frac{d\langle p\rangle}{dt} &= -\langle\frac{dV}{dx}\rangle= 2(\langle e^{-2x}\rangle-\langle e^{-x}\rangle)\\
\frac{d\langle p^2\rangle}{dt} &= -\langle p\frac{dV}{dx}+\frac{dV}{dx}p\rangle- \sqrt{2\pi}\gamma\frac{d}{dt}\int|\psi|^4dx\\
\frac{d\langle xp+px\rangle}{dt} &=2\langle p^2\rangle-2aD\langle x\frac{dV}{dx}\rangle+\sqrt{2\pi}\gamma\int|\psi|^4dx\\
&= 2\langle p^2\rangle+4aD(\langle xe^{-2x}\rangle-\langle xe^{-x}\rangle)\\&+\sqrt{2\pi}\gamma\int|\psi|^4dx
\end{aligned}
\end{eqnarray}
With the help of the equations given in Eq. (\ref{eq13}), we find that the dynamics of the wave packet will be governed by the two coupled equations given below
\begin{eqnarray}
\frac{d^2}{dt^2}x_0=2[e^{-(2x_0-\Delta^2)}-e^{-(x_0-\frac{\Delta^2}{4})}]\label{eq14}
\end{eqnarray}
\begin{eqnarray}\label{eq15}
\begin{aligned}
\frac{d^2}{dt^2}\Delta^2=\frac{2}{K^2}\frac{1}{\Delta^2}+\frac{\gamma}{\Delta}+4\Delta^2[&\frac{1}{2}e^{-(x_0-\frac{\Delta^2}{4})}\\&-e^{-(2x_0-\Delta^2)}]
\end{aligned}
\end{eqnarray}
Here we have explored the wave packet dynamics for  $K=2$ which fixes $\gamma_c=0.917$. 
\subsection{\bf{Dynamics: ($\gamma>0$ and $\gamma<\gamma_c$); Corresponding figure: Fig \ref{fig2}}} 
In this parameter region after solving the coupled equations given in Eq. (\ref{eq14}) and Eq. (\ref{eq15}) numerically, we have found that when initial momentum ($p_0$) of the wave packet is small, it reflects back from the potential barrier at right. Then it moves to the left and collides with the infinite barrier or the potential at right. This back and forth oscillation within the trap continues with time. With the increase of $p_0$, the oscillation of the peak of the wave packet within the trap goes on with the higher amplitude up to a certain threshold value ($(p_{th}$) of $p_0$. At $p_0=p_{th}$ it simply comes out of the potential barrier and hence the oscillatory behavior of $x_0(t)$ ceases. For $p_0>p_{t})$ the graph for $x_0(t$) vs. $t$ shows sharp linear increase (not shown in given figures). For $K=2$, we observed $p_{th}=0.45$. Hence with the help of Eq. (\ref{a}), we conclude that the initial wave packet has to have the minimum average energy $E=0.751D$ to come out of the potential barrier. We will call this minimum average energy required for emitting the wave packet from the potential as the threshold energy ($E_{th}$). It should be noted that in the classical case, for a partice to escape from the Morse potential, the average $E$ needs to be greater than or equal to zero. If the particle is at the origin initially with a potential energy of -1, then the threshhold momentum would be $p_0=\sqrt{2}$. In the quantum case the average initial momentum is 0.45 clearly showing the role of quantum fluctuations. The dynamics of the width also follows the same qualitative feature as described above. The dynamics in this parameter region is clearly depicted in Fig \ref{fig2}. For convenience we keep the initial width of the wave packet fixed at $\Delta_0=0.4$ throughout the study.

\begin{figure}[H]
\includegraphics[angle=0,scale=0.73]{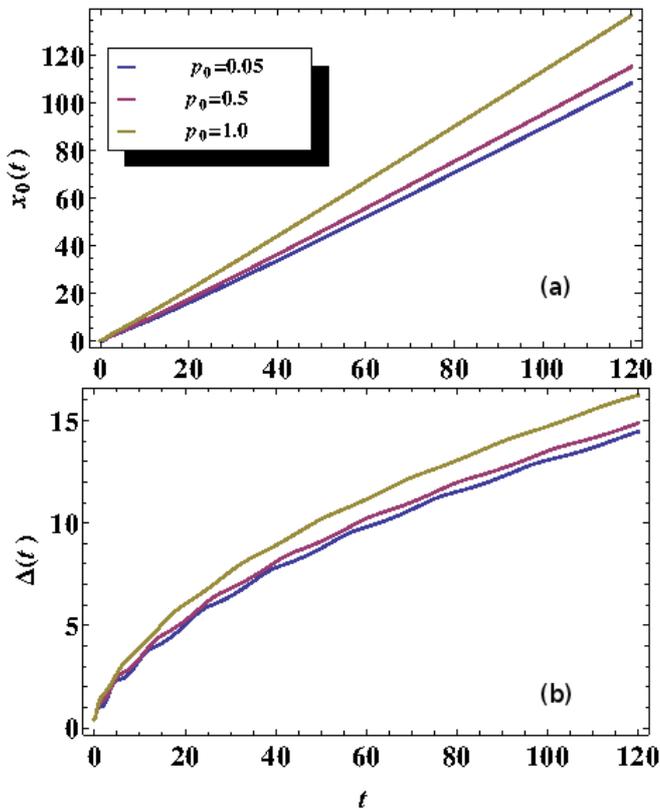}
\caption{Figure (a) shows the dynamics of the peak ($x_0$) of the wave packet for $\gamma=1.2$ for three different values of $p_0$. The dynamics of $x_0$ always shows linear increase for all these three values of $p_0$ unlike the previous case of $\gamma<\gamma_c$. This implies in this parameter region the wave packet will always come out of the potential barrier whatever be the initial momentum. The width also increases and the dynamics is like that of a free particle.}
\label{fig3}
\end{figure}

\subsection{\bf{Dynamics: ($\gamma>0$ and $\gamma>\gamma_c$); Corresponding figure: Fig \ref{fig3}}} 
\begin{figure}[H]
\includegraphics[angle=0,scale=0.73]{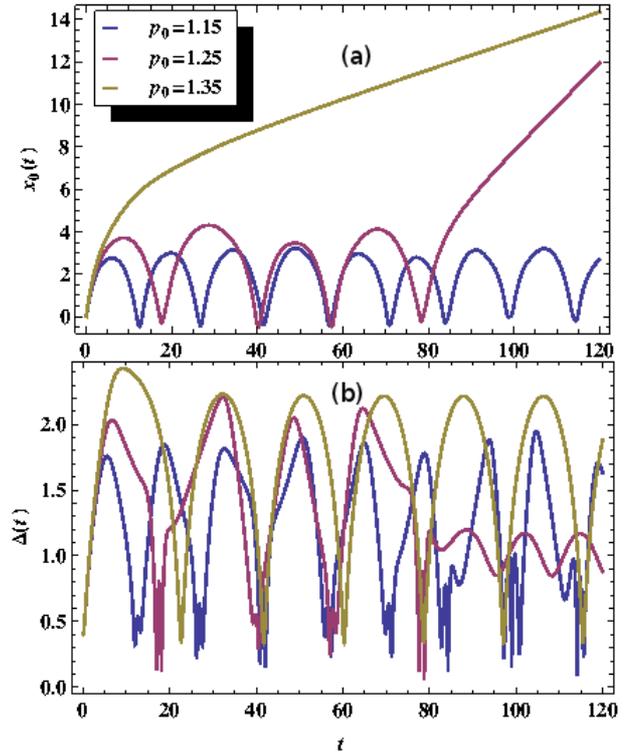}
\caption{Figure (a) shows the dynamics of the peak ($x_0$) of the wave packet for $\gamma=-0.5$ for three different values of $p_0$. $K=2$ and ($\Delta_0=0.4$) are considered. Like Figure 2(a), with lower values of $p_0$ the wave packet hits the potential barrier and reflects back and forth within the trap. $x_0(t)$ shows oscillatory behavior in this case also. With increase of $p_0$ the number of reflection reduces, oscillatory behavior ceases and finally when $p_0=p_{th}(=1.35)$ the wave packet comes out of the potential barrier indicated by the gradual increase of $x_0(t)$. In figure (b), the dynamics of width shows a random oscillation for $p_0<p_{th}(=1.35)$ and as $p_0\rightarrow p_{th}$, the oscillation takes more periodic manner with constant amplitude. But unlike the previous cases here the width is bounded by a maximum and minimum value ($\Delta_0$).}
\label{fig4}
\end{figure}

\begin{figure}[H]
\includegraphics[angle=0,scale=0.7]{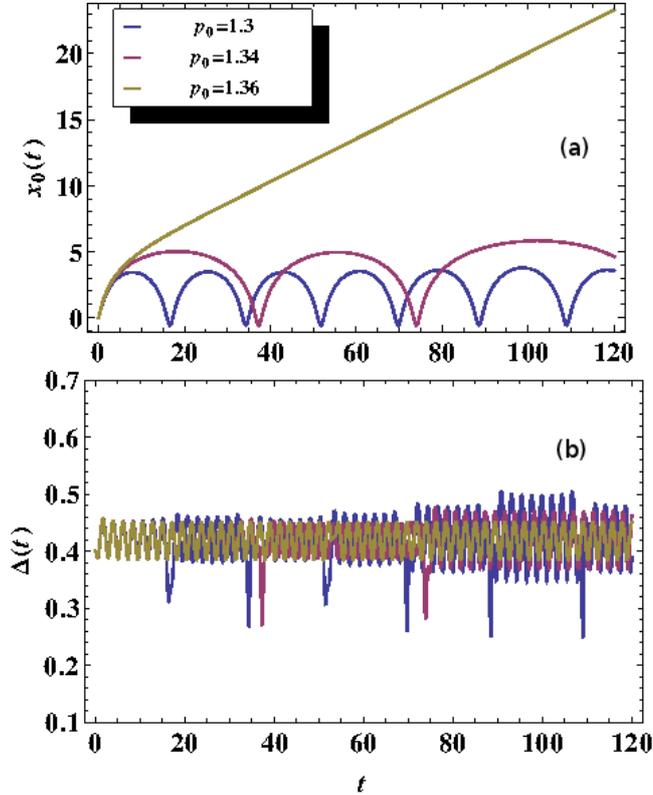}
\caption{Figure (a) shows the dynamics of the peak ($x_0$) of the wave packet for $\gamma=-1.2$ for three different values of $p_0$. $K=2$ and ($\Delta_0=0.4$) are considered. Like Fig. 4(a), with low values of $p_0$ the wave packet hits the potential barrier and reflects back and forth within the trap, showing oscillatory behavior. With increase of $p_0$ the number of reflection reduces, oscillatory behavior ceases and finally when $p_0=p_{th}(=1.35)$ the wave packet comes out of the potential barrier indicated by the gradual increase of $x_0(t)$. In figure (b), the dynamics of width shows periodic oscillation for all values of $p_0$ and as $p_0\rightarrow p_{th}$, the amplitude of oscillation reduces. At $p_0=p_{th}$, the width remains almost constant for all time.}

In Table 2., we have calculated and compared the values of threshold energy for different values of $K$. For $\gamma>0$, as $K$ increases the value of $E_{th}$ increases which is justified by the fact that with increase of $K$, Morse potential approaches oscillator limit (more highly trapped limit) and consequently greater energy is required for escaping from the potential. 
\label{fig5}
\end{figure}

In this parameter region we have observed the linear increase of $x_0(t)$ irrespective of initial momentum ($p_0$) of the wave packet. The width also shows continuous increase with time. With the help of Eq. (\ref{a}), we now conclude that wave packet in this parameter region always have energy $E> E_{th}$($=0.751D$ obtained for $K=2$ and $\gamma=0.5$). Thus the wave packet always comes out of the potential barrier and delocalises in space. In Fig \ref{fig3}, we clearly describe this feature considering $\gamma(=1.2)>\gamma_c(=0.917)$, $\Delta_0=0.4$ and $K=2$. The behavior is always thatof a free particle.
\subsection{\bf{Dynamics{($\gamma<0$); Corresponding figure: Fig \ref{fig4} and Fig\ref{fig5}}}}
In this parameter region the dynamics of $x_0$ shows the same qualitative feature as is described in Figure \ref{fig2}. We set $\gamma=-0.5$, $K=2$ and $\Delta_0=0.4$. But unlike the previous cases now the dynamics of the width is bounded. The random oscillation of width continues with time with moderate amplitude and as $p_0\rightarrow p_{th}$ the oscillation takes place in a more periodic manner with constant amplitude which implies that the wave packet will never delocalise in space after coming out of the potential barrier. This case is depicted in Fig \ref{fig4}. In Fig \ref{fig5} we have shown the dynamics of $x_0$ and $\Delta$ for a greater value of $\gamma=-1.2$. The dynamics of $x_0$ follows the same qualitative feature as described earlier. But this time the dynamics of the width shows an interesting feature. The width of the wave packet has a tendency to retain its initial shape within the trap as well as after coming out of the trap.
\begin{table}[H]
 \centering 
\begin{tabular}{c  c  c} 
\hline\hline 
$K$ & $E_{th}$(for $\gamma=0.5$) & $E_{th}$(for $\gamma=-0.5$)  \\ [0.5ex]  
\hline 
2 & 0.751D & -0.934D\\
3 & 0.904D & -1.09D\\
4 & 0.971D & -1.14D\\
5 & 0.986D & -1.17D\\ 
6 & 1.004D & -1.185D\\[1ex]
\hline 
\end{tabular}
\caption{Results showing how the threshold energy ($E_{th}$) behaves with potential paramater $K$ for two fixed values of $\gamma$ ($\gamma=0.5$ and $\gamma=-0.5$)}
\end{table}
\section{Conclusion}
In conclusion, we have explored the features of the GPE with the Morse potential. We have found that for a positive coupling constant and a deep potential there is a critical coupling $g_c$ where the ground state disappears as a bound state and $g_c^{4/3}$ scales as the depth of the potential. For the dynamics in this potential if $g<g_c$, the initial wave packet needs to posses a threshold average momentum for the packet to escape from the potential. For $g>g_c$ however, the wave packet dynamics always resembles that for a quantum free particle. If the coupling $g$ is negative, we find that the packet does escape from the potential for above threshold value of the average momentum and further if $g$ is more negative than a critical value, the width of the packet remains constant in time.  

\section*{Acknowledgments}
One of the authors, Sukla Pal would like to thank S. N. Bose National Centre for Basic Sciences for the financial support during the work. Sukla Pal acknowledges Harish-Chandra Research Institute for hospitality and support during visit.

\end{document}